\documentclass[letterpaper]{IEEEtran}
\usepackage[letterpaper, left=0.75in, right=0.75in, bottom=1.2in, top=1in]{geometry}

\usepackage{cite}
\usepackage{amssymb,amsmath,times,wasysym,comment,amsfonts}
\usepackage{algorithmic}
\usepackage{psfrag,graphicx}
\usepackage{textcomp,comment}
\usepackage{booktabs}

\title{Joint Beam Hopping and Carrier Aggregation in High Throughput Multi-Beam Satellite Systems\vspace{2mm}}
	
\author{\IEEEauthorblockN{Mirza Golam Kibria\IEEEauthorrefmark{1},~\IEEEmembership{Member,~IEEE},
			Hayder Al-Hraishawi\IEEEauthorrefmark{2},~\IEEEmembership{Member,~IEEE},\\
			Eva Lagunas\IEEEauthorrefmark{2},~\IEEEmembership{Senior Member,~IEEE},
			Symeon Chatzinotas\IEEEauthorrefmark{2},~\IEEEmembership{Senior Member,~IEEE},\\ and
			Bj\"orn Ottersten\IEEEauthorrefmark{2},~\IEEEmembership{Fellow,~IEEE},\\
			\IEEEauthorrefmark{1}Department of Research and Development, Huawei, Stockholm, Sweden.}\\
			\IEEEauthorrefmark{2}Interdisciplinary Centre for Security, Reliability and Trust (SnT), University of Luxembourg. \vspace{-1mm}
			
			\thanks{This research was funded in whole by the Luxembourg National Research Fund (FNR) in the frameworks of the FNR projects ``MegaLEO: Self-Organised Lower Earth Orbit Mega-Constellations''  (Grant no. C20/IS/14767486) and ``DISBuS: Dynamic Beam Forming and In-band Signalling for Next Generation Satellite Systems'' (BRIDGES19/IS/13778945/DISBuS).\\ For the purpose of open access, the authors have applied a Creative Commons Attribution 4.0 International (CC BY 4.0) license to any Author Accepted Manuscript version arising from this submission}
	}

\begin{document}

\maketitle

\thispagestyle{plain}
\pagestyle{plain}

\begin{abstract}
	Beam hopping (BH) and carrier aggregation (CA) are two promising technologies for the next generation satellite communication systems to achieve several orders of magnitude increase in system capacity and to significantly improve the spectral efficiency. While BH allows a great flexibility in adapting the offered capacity to the heterogeneous demand, CA further enhances the user quality-of-service (QoS) by allowing it to pool resources from multiple adjacent beams. In this paper, we consider a multi-beam high throughput satellite (HTS) system that employs BH in conjunction with CA to capitalize on the mutual interplay between both techniques. Particularly, an innovative joint BH-CA scheme is proposed and analyzed in this work to utilize their individual competencies. This includes designing  an efficient joint time-space beam illumination pattern for BH and multi-user aggregation strategy for CA. Through this,  user-carrier assignment, transponder filling-rates, beams hopping pattern, and illumination duration are all simultaneously optimized by formulating a joint optimization problem as a multi-objective mixed integer linear programming problem (MINLP). Simulation	results are provided to corroborate our analysis, demonstrate the design	tradeoffs, and point out the potentials of the proposed joint BH-CA concept. Advantages of our BH-CA scheme versus the conventional BH method without employing CA are investigated and presented under the same system circumstances.
\end{abstract}
	

\section{Introduction}
	Satellite communications technologies are currently experiencing a remarkable evolution due to a paradigm shift brought about by the development of software-defined satellites \cite{Pan2015}. The latest satellite technologies have evolved traditional satellite payloads into increasingly digitally reconfigurable versions in order to offer generic and software-based solutions as well as to accommodate spatially and temporally diverse and unpredictable demand \cite{NGSO_surevy}. Moreover, satellite systems have unequivocal advantages such as providing ubiquitous coverage over vast geographies, wideband transmissions, navigation services, and mission-critical operations \cite{Sharma2018}. Thus, the demand for satellite services is rapidly growing to provide affordable, accessible, uninterrupted wireless connectivity especially to  underserved and unserved areas. Beyond  traditional satellite applications like aeronautical, maritime, mapping, weather forecasting, broadcasting, rescue and disaster relief, recent advances in satellite technologies, especially within the non-geostationary orbits (NGSO), have unlocked the potential of satellites to carry and execute innovative applications and new services from space \cite{SIN_2021}. 
	
	Naturally, satellite traffic is highly diversified and spatially distributed over the coverage areas from various users with variable quality-of-service (QoS) profiles, which inflicts daunting operational challenges for accommodating the asymmetry and heterogeneity of the traffic demands \cite{Tang2021}. Besides, satellite resources are very scarce and not always affordable particularly the radio frequency (RF) spectrum and the RF chains \cite{Colin2022}. 
	Hence, conventional resource allocation schemes mapping spectrum and payload power to satellite beams suffer from capacity deficiency in hot-spots and under-utilization in so-called cold-beams \cite{Shahid2020}. Therefore, it is imperative to devise innovative and efficient techniques exploiting the potential of software configured satellites to improve resource utilization while satisfying the burgeoning traffic demand and the high throughput requirements \cite{Hayder2021icc}. 
	
	In this direction, many  contributions have developed different resource allocation techniques to utilize satellite resources in an efficient manner. For instance, a flexible frequency allocation method is proposed in \cite{Kawamoto2020} to mitigate the inter-beam interference issues in satellite communication systems. With the goal of improving the flexibility of resource allocation  to satisfy the geographic distribution of traffic requirements, the authors of \cite{Takahashi2022} have designed a power resource allocation model with a digital beamforming (DBF)-based fusion control in the high-throughput satellite (HTS) systems. In \cite{Roumeliotis2021}, optimal dynamic capacity allocation schemes have been proposed in a multi-beam smart gateway diversity HTS system with non-regenerative satellites in order to minimize system’s capacity losses as well as to satisfy various quality-of-service (QoS) requirements of the served users. Furthermore, an algorithm to efficiently distribute the inter-orbit satellite resources in the non-terrestrial networks (NTNs) is  developed in \cite{Dazhi2022} with taking into consideration uplink data rate and communication latency.

	Furthermore, an intriguing concept was recently introduced to multi-beam HTSs, that is beam hopping (BH),  in order to  provide the required high flexibility for accommodating irregular and time varying traffic demands as well as to decrease system power consumption  \cite{Hu2022,Lin2022}.  
	In BH-based satellite systems, the serving beams are activated/illuminated using time division where only a subset of the available beams is activated at a time instance according to their traffic volumes. The advantages of employing the BH scheme in multi-beam HTS systems are manifold. In addition to the high flexibility in resource allocation in both temporal and spacial domains, it allows scaling down the payload mass by reducing the required number of RF chains. This is highly beneficial and  will definitely alleviate the strong cost constraints on satellite launching and operation.
	In essence, the BH scheme offers higher allocation flexibility, and consequently, improves the resource utilization and increases the achievable system capacity. Thereby, the BH concept has gained momentum recently in both academia and industry where, for example, several research projects were funded by the European Space Agency (ESA) to investigate the potential and use cases of BH technique \cite{ESA_BH,ESA_BEHOP,ESA_FlexPreDem}. 
	
	Along with introducing BH as a promising technology to enhance resource utilization, there has been a prominent solution to further increase spectral efficiency and the overall system capacity through carrier aggregation (CA) \cite{Kibria2019,Hayder2020}. CA is a successfully implemented technology in terrestrial networks, which allows network operators to aggregate multiple component carriers across the available spectrum to achieve higher peak data rates \cite{Rui2013}. Correspondingly, CA can be employed within the HTS architectures to ensure a high-quality user experience and to harness the multiplexing gain via flexibly distributing traffic demands over multiple carriers \cite{Lagunas2020}. Specifically, in satellite communication systems, each transponder operates over a certain spectrum and an HTS transponder can amplify single or multiple carriers, in this context CA can balance the traffic loads among the serving beams that experience  heterogeneous demands. Motivated by these benefits, a few research works have been conducted in the satellite communication domain to explore CA technical feasibility and its potential returns \cite{ESA_CADSAT,CA_Mirza,Hayder2021a}, where it has been concluded that CA does not only circumvent the spectrum scarcity issue but it also  ameliorates user-fairness in terms of QoS and throughput.

	Individually, each of BH and CA has the potential of  offering an order of magnitude or more increase in system throughput compared to the conventional techniques. Fortunately, these two solutions share a symbiotic convergence in many standpoints: BH scheme adapts the offered capacity to spatial fluctuations the demand  in a flexible fashion, while CA schemes can ensure high QoS for users through aggregating resources from multiple adjacent beams along with boosting system capacity by performing redistribution (optimization) of the component carriers from a single or multiple transponders. Indeed, harnessing the synergetic interaction between these two approaches would yield an improved system performance in many aspects and certainly provide high-speed transmission data rates. Taking a step in this direction, we propose an innovative approach to combine BH and CA, that is joint BH-CA, to benefit from the properties of each one. In particular, the BH scheme brings forth flexible satellite resource utilization in the time domain whereas the CA scheme increases the flexibility in the frequency domain. 
	
	In this work, we focus on studying the feasibility of the proposed concept by jointly devising the time-space transmission plan and user-carrier assignment along with the carrier filling-rates. Thereby, the key technical contributions in this article can be summarized as follows:
	\begin{itemize}
		\item Developing a novel BH technique in synergy with CA concept for HTS systems to construct a joint BH-CA resource allocation scheme benefiting from the intrinsic features of both technologies. To the best of our knowledge, the joint BH-CA interaction has not yet been investigated in the open literature.
		
		\item In the proposed combination, multiple parameters and variables need to be concurrently considered and optimized such as traffic demand, channel conditions, user-carrier allocation, beams illumination pattern and ratio; therefore, a multi-objective MINLP is formulated and solved for this purpose.  
		
		\item  The performance of the proposed BH-CA concept is investigated herein in terms of satisfying the traffic demand and rate matching. Simulation results including  performance comparisons are provided to demonstrate the validity and gains of the proposed technique.
		
	\end{itemize}
	
	The reminder of this article is organized as follows. The system model is presented in Section \ref{sec:system_model} and the problem formulation is stated in Section \ref{sec:problem_formulation}. The proposed solution for the optimization problem is provided in Section \ref{sec:proposed_solution}. In Section \ref{sec:simulation_results}, the simulation results are provided to evaluate the proposed approach. This article is then concluded in Section \ref{sec:conclusions}.  We list the symbols that will be frequently used in this paper along with their definitions in Table \ref{tab:symbols} for ease of reference.
	
	\begin{table}[t] \centering
		\caption{List of used symbols} \label{tab:symbols}
		\begin{tabular}{|l|l|}
			\hline
			Symbol              & Description  									\\ \hline
			$N_{\rm B}$  & Number of the serving beams  		\\ \hline
			$L$ 		 & Number of clusters in the system	\\ \hline
			$T_{\rm H}$ & Hopping window duration 	\\ \hline
			$N_{\rm TS}$&  Number of time-slots in hopping windows \\ \hline
			$T_{\rm slot}$ & Time-granularity of the BH operation \\ \hline
			$N_{\rm T}$  & 	Number of active cluster in a time-slot 	\\ \hline
			$\mathcal{R}^{(l)}_{c,u}$     & \begin{tabular}[c]{@{}l@{}} Achievable data rate of the user $u$ under cluster $l$ \\ when it is assigned to carrier $c$\end{tabular}                          \\ \hline
			$\gamma^{(l)}_{c,u}$ & Signal-to-interference-plus-noise ratio  \\ \hline
			$f_{\rm SE}$ & Spectral efficiency mapping function  \\ \hline
			$\Delta_{\rm max}$  & Maximum  data carriers  \\ \hline
			$\beta^{(l)}_{c,u}$ & \begin{tabular}[c]{@{}l@{}}Fractions of component carrier $c$ assigned to user $u$ \\ under cluster $l$\end{tabular} \\ \hline
			$z_{l,t}$           & \begin{tabular}[c]{@{}l@{}}Binary variable determines if cluster $l$ is active \\ in time-slot $t$ or not ($z_{l,t}=1$:active, $z_{l,t}=0$:inactive)\end{tabular}                           \\ \hline
			$a^{(l)}_{c,u}$     & Carrier-user assignment indicator\\ \hline
			$\mathcal{U}^{(l)}$ & Set of users within the $l$-th cluster  		\\ \hline
			$\mathcal{C}^{(l)}$ & Set of carriers used in the $l$-th cluster 	\\ \hline
			$\mathcal{B}^{(l)}$ & Set of beams assigned to the $l$-th cluster 	\\ \hline
			$s_{u,l}$ & Offered capacity to user $u$ that belongs to the $l$-th cluster \\ \hline
			$s_{l}$ & Total offered capacity to cluster $l$ \\ \hline
			$d_{u,l}$ & Demand of user $u$ within the coverage of the $l$-th cluster\\ \hline 
			$ d_{l}$ &  Total demand of $l$-th cluster \\ \hline 
		\end{tabular}
	\end{table}

	\begin{figure*}
		\centering
		\includegraphics[scale=0.600]{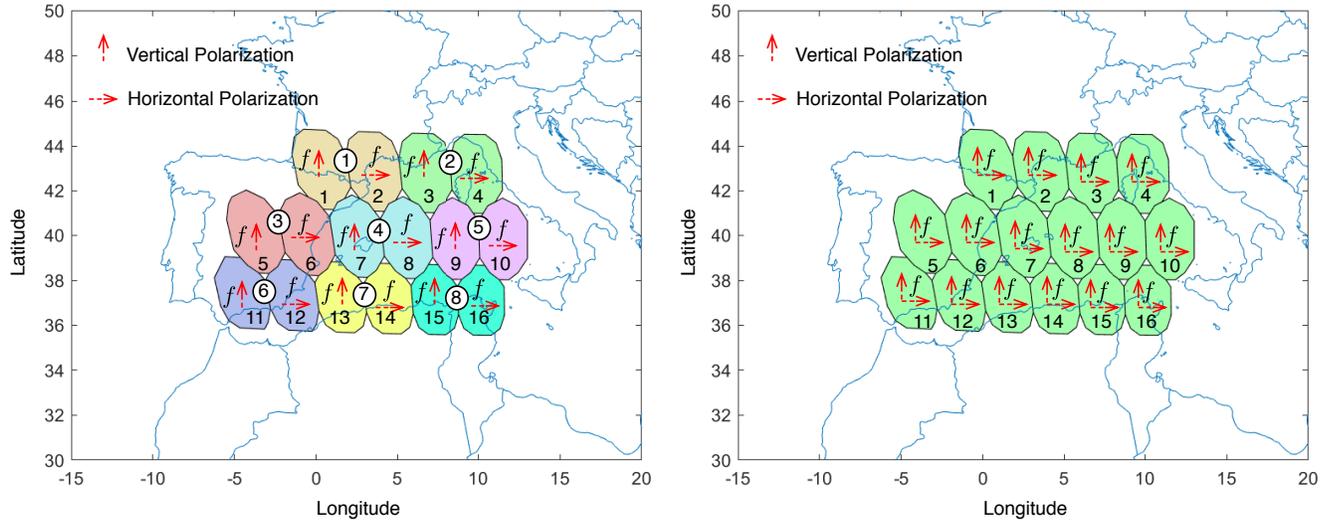}
		\caption{Left: The satellite beam-pattern with 16 beams along with the frequency and polarization allocation plan for a joint BH-CA based satellite system. Beams with the same colors but with opposite polarization form the clusters. We employ a non-overlapping clustering process where one beam can be part of only one cluster. Each beam is served with the whole system bandwidth but in a single polarization. Right: Frequency and polarization allocation plan for conventional beam-hopping system. Each beam is served with two different polarization. The transponder bandwidth is the satellite system bandwidth.}
		\label{fig_BHCA_y}
		\psfrag{Demand}{$Demand$}
	\end{figure*}

	\section{System Model}\label{sec:system_model}
	We consider a multi-beam dual-polarization HTS system  with a total of $N_{\rm B}$ beams as depicted in the beam coverage pattern in Fig.~\ref{fig_BHCA_y}. The  coverage area of the $N_{\rm B}$ beams is divided into  clusters ($N_l$) for $l \in \{1,\cdots L\}$, where each cluster consists of two adjacent beams and each beam serves with the full available satellite bandwidth because the beams in the clusters operate in an orthogonal polarization. This is also known as cluster hopping, in which,  multiple set of adjacent beams are illuminated at the same time with the same frequency resource \cite{lagunas2021}. In this setting, multiple users can be multiplexed within a single carrier. Further, when CA is employed, the user data can be delivered through two component carriers of opposite polarization.
	
	In BH operation mode, the satellite resources in this mode, i.e. transmit power and transponders, are determined based on the average demand (not the peak demand). The number of available transponders as well as the number of beams in each cluster determine the so-called illumination ratio. Specifically, the illumination ratio represents the number of illuminated beams at a time instance to the number of available beams in the system. In this context, there is no exclusive transponder\footnote{By transponder, we mean an RF chain consisting of a high power amplifier and other additional components such as filter, low-noise amplifier, multiplexer, etc., handling the signal transmission over a given polarization. The transponders can have just one component carrier or multiple carriers.} for any beam and the transponders are shared among all the serving beams and they are associated with activated beams periodically based on the illumination pattern. Namely, only a subset of clusters/beams are illuminated simultaneously, which requires a less number of transponders (i.e. high power amplifiers (HPAs)) than the number of beams. Under the BH scheme, the service time is divided into hopping windows with duration of $T_{\rm H}$ that repeats in a cyclic way. Each hopping window consists of $N_{\rm TS}$ time-slots where the duration of a time-slot, $T_{\rm slot}$ accounts for the time-granularity of the BH operation. In this system, we consider that at each time-slot only $N_{\rm T}$ ($N_{\rm T}< L$) clusters are active, Thus, the users belonging to the active clusters are served and each cluster has two data carriers with different polarization to serve the users.

	In this setup, the achievable data rate (supply capacity) of the user $u$ under cluster $l$ when it is assigned to carrier $c$ in a certain time-slot is denoted by $\mathcal{R}^{(l)}_{c,u}$ and  can be computed based on the channel-state information (CSI) and the link budget as
	\begin{equation}
		\mathcal{R}^{(l)}_{c,u}=B \left( f_{\rm SE}(\gamma^{(l)}_{c,u})\right),
	\end{equation}
	where $B$ denotes the carrier bandwidth in Hz, and $\gamma^{(l)}_{c,u}$ accounts for the signal-to-interference-plus-noise ratio
	(SINR) for the $u$-th user in the $l$-th cluster served by the $c$-th carrier. The function $f_{\rm SE}$ is the spectral efficiency (SE) mapping function according to the adaptive coding and modulation (ACM) scheme considered by the digital video broadcasting-satellite (DVB-S2X) specifications \cite{DVBS2X}. 
	
	For the CA scheme, users can be granted with a maximal access to the available data carriers that is defined as $\Delta_{\rm max}$ or a partial of  $\Delta_{\rm max}$ based on the fill-rate factor ($\beta^{(l)}_{c,u}$) that determines the fractions of data carrier $c$ assigned to user $u$ under the $l$-th cluster. Furthermore, a binary parameter ($z_{l,t}$) is introduced to indicate whether cluster $l$ is activated in time-slot $t$ or not ($z_{l,t}=1$ $\rightarrow$ active, $z_{l,t}=0$ $\rightarrow$ inactive). Additionally, we need to define a carrier-user binary assignment variable ($a^{(l)}_{c,u}$), namely $a^{(l)}_{c,u} = 1$ when carrier $c$ of cluster $l$ is assigned to user $u$, and  otherwise $a^{(l)}_{c,u} = 0$. Based on these considerations, the offered capacity to user $u$ that belongs to cluster $l$ can be calculated as
	\begin{equation}
		s_{u,l}=\sum_{t=1}^{N_{\rm TS}}\sum_{\forall c, c\in \mathcal{C}^{(l)}}\beta^{(l)}_{c,u} \mathcal{R}^{(l)}_{c,u}z_{l,t},\forall  u,u\in\mathcal{U}^{(l)},
	\end{equation}
	where $\mathcal{U}^{(l)}$ is the set of users belong to cluster $l$ and $\mathcal{C}^{(l)}$ is the set of carriers in cluster $l$. 
	Thereby, the total offered capacity to cluster $l$ is given by
	\begin{equation}
		s_{l}=\sum_{\forall u, u\in\mathcal{U}^{(l)}}s_{u,l},\hspace{1mm}\forall l, l\in L.
	\end{equation}

	\section{Problem Formulation}\label{sec:problem_formulation}
	
	In our system, the offered capacity has to be tailored to the user demands within the clusters, particularly let $d_{u,l}$ be the demand of user $u$ within the coverage of cluster $l$ while $ d_{l}=\sum_{u\in\mathcal{U}^{(l)}}d_{u,l}$ be the total demand in the $l$-th  cluster.
	The objectives here are two-fold; the parameters $\beta^{(l)}_{c,u}$, $a^{(l)}_{c,u}$ and $z_{l,t}$ need to be optimized for each cluster such that the minimum of user supply to demand ratio $\frac{s_{u,l}}{d_{u,l}},u \in \{1,\cdots N^{(l)}_{\rm U}\}$ is maximized, where $N^{(l)}_{\rm U}$ is the number of users under cluster $l$.
	At the same time we want to maximize the minimum of $\frac{s_{l}}{d_{l}},l \in \{1,\cdots L\}$.
	In particular, we opt to maximize the minimum offered capacity to traffic ratio user-wise and cluster-wise at the same time. Accordingly, the joint BH-CA problem formulation is given as 
	\begin{equation}
		\label{main:ps1}
		\begin{array}{*{35}{l}}
			\underset{a^{(l)}_{c,u},\beta^{(l)}_{c,u},z_{l,t}}{\max}\hspace{1.5mm}\underset{u,u=1,\cdots,N^{(l)}_{\rm U}}{\min}\hspace{1mm}\frac{s_{u,l}}{d_{u,l}},\forall l,\underset{a^{(l)}_{c,u},\beta^{(l)}_{c,u},z_{l,t}}{\max}\hspace{1mm}\underset{l=1,\cdots,N_{\rm L}}{\min}\hspace{1mm}\frac{s_{l}}{d_{l}}\vspace{2mm}  \\ 
			
			\text{}\text{subject to }\text{ C1:} \hspace{2mm}\sum\limits_{\forall c, c\in \mathcal{C}^{(l)}} a^{(l)}_{c,u}\le \Delta_{\max},\forall  u,u\in\mathcal{U}^{(l)}, \vspace{2mm} \\
			\text{}\hspace{15mm}\text{ C2:} \hspace{2mm}\sum\limits_{u=1}^{N^{(l)}_{\rm U}} \beta^{(l)}_{c,u}\le 1, \forall  c,c\in\mathcal{C}^{(l)}, \vspace{2mm} \\
			\text{}\hspace{15mm}\text{ C3:} \hspace{2mm}\sum\limits_{l=1}^{N_{\rm L}}z_{l,t}\le N_{\rm T},\forall t, t\in \{1,\cdots, N_{\rm TS}\}\\
		\end{array}
	\end{equation}
	
\begin{figure}
		\centering
		\includegraphics[scale=0.54]{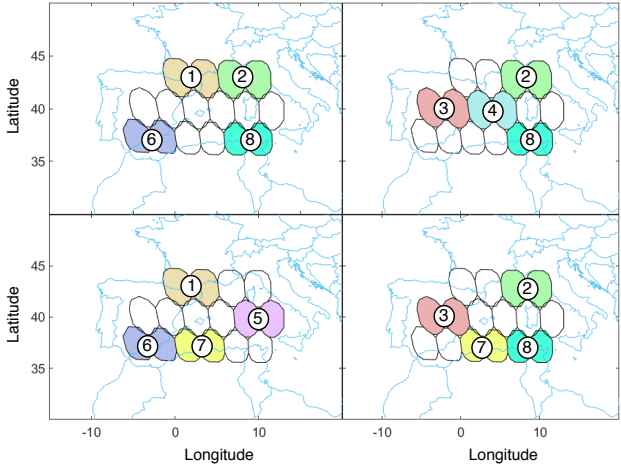}
		\caption{Depiction of some valid BH snapshots for the satellite beam pattern in Fig.~\ref{fig_BHCA_y} for $\frac{1}{4 }$ illumination ratio where each snapshot contains two non-adjacent clusters. }
		\label{fig_CADSAT_res6}
		\psfrag{Demand}{$Demand$}
	\end{figure}

	Basically, user-carrier assignment and the corresponding fill-rates for the users in a certain cluster remain the same throughout the hopping window. This feature of the proposed joint BH-CA scheme is consistent with the cluster hopping technique for non-overlapping clusters \cite{lagunas2021}, i.e. a beam cannot be a part of multiple clusters. With the overlapping clustering approach, any group of adjacent beams with opposite polarization can form a cluster, see Fig.~\ref{fig_CADSAT_res6} for an illustration.
	
	The optimization problem in  \eqref{main:ps1}  is defined for the hopping window duration that repeats over time. Therefore, the demand and offered capacity of the users that are in bits per second (bps) will be scaled down to bphw (bits per hopping window). It is straight forward to eliminate the $\max-\min$ formulation above by introducing some slack variables such as $t^{(l)}_{\rm U},l=1,2,{N_{\rm L}}$ and $t_{\rm L}$ and imposing the following additional constraints
	\begin{equation}
		\begin{array}{*{35}{l}}
			\text{C4: }s_{u,l}\ge t^{(l)}_{\rm U} d_{u,l},l=1,2,\cdots,{L} \vspace{2mm} \\
			\text{C5: }s_{l}\ge t_{\rm L} d_{l}, l\in L
		\end{array}
	\end{equation}
	
	It is important to mention that with higher frequency re-use factor, overall shape of the satellite beam pattern and clustering process, optimal polarization planning (when allowing adjacent beams to operate using same polarization) may not be possible, which eventually brings about high inter-cluster interference. Furthermore, insufficient beam separation or isolation would also increase the inter-beam interference due to the coverage overlapping with the adjacent beams. 
	The best way to deal with such practical considerations is to inflict a constraint to prevent adjacent clusters from a simultaneous illumination and the clustered beams must have at least a layer of separation. For example, polarization is a good isolation feature to consider where the illuminated beams operate on opposite polarization in order to minimize the inter-cluster interference. To this end, we can enforce the simultaneously illuminated clusters to be non-adjacent by adding the following constraint.
	\begin{equation}
		\text{C6: } z_{n_1,t} + z_{n_2,t} \leq 1, \forall (n_1, n_2)\in \mathcal{P} \text{ with } \mathcal{P}_{n_1 n_2}=1
	\end{equation}
	where $\mathcal{P}$ is the set of all adjacent pairs of clusters without a layer of separation. The set $\mathcal{P}$ can be easily calculated from the adjacency matrix of the beams/clusters and polarization planning.
	
	The constraints in the considered problem formulation in \eqref{main:ps1} do not really provide the expected mapping between the association indicator $a^{(l)}_{c,u}$ and the fill-rate variable $\beta^{(l)}_{c,u}$. We expect that after solving the problem, if the output $\beta^{(l)}_{c,u} >0$, we should have $a^{(l)}_{c,u}=1$ or vice versa, but this may not be the case as we have not defined any relation between the variables $a^{(l)}_{c,u}$ and $\beta^{(l)}_{c,u}$. Moreover, since the constraint C4 is an important design aspect for CA, there should be one-to-one mapping between $a^{(l)}_{c,u}$ and $\beta^{(l)}_{c,u}$.  This issue can be addressed by introducing the following {\textit{conditional}} constraints to the existing problem formulation in \eqref{main:ps1}.
	
	\begin{equation}
		\label{eqcc}
		\text{C7}: \left\{ \begin{array}{*{35}{l}}
			\text{N1}: \beta^{(l)}_{c,u}=0 \;\; \text{ if }a^{(l)}_{c,u}=0\\
			\text{N2}: \beta^{(l)}_{c,u}>0 \;\; \text{ if }a^{(l)}_{c,u}=1\\
		\end{array}\right.
	\end{equation}

	Thus, the optimization problem with considering the added constraints can be rewritten as
	\begin{equation}
		\label{main:ps2}
		\begin{array}{*{35}{l}}
			\underset{a^{(l)}_{c,u},\beta^{(l)}_{c,u},z_{l,t}}{\max}\hspace{1mm}t^{(1)}_{\rm U},\cdots,\underset{a^{(l)}_{c,u},\beta^{(l)}_{c,u},z_{l,t}}{\max}\hspace{1mm}t^{(N_{\rm L})}_{\rm U},\underset{a^{(l)}_{c,u},\beta^{(l)}_{c,u},z_{l,t}}{\max}\hspace{1mm}t_{\rm L}\\
			\text{}\text{subject to }\text{ C1:} \hspace{2mm}\sum_{\forall c, c\in \mathcal{C}^{(l)}} a^{(l)}_{c,u}\le \Delta_{\max},\forall  u,u\in\mathcal{U}^{(l)}, \vspace{1.5mm} \\
			\text{}\hspace{15mm}\text{ C2:} \hspace{2mm}\sum_{u=1}^{N^{(l)}_{\rm U}} \beta^{(l)}_{c,u}\le 1, \forall  c,c\in\mathcal{C}^{(l)}, \vspace{1.5mm} \\
			\text{}\hspace{15mm}\text{ C3:} \hspace{2mm}\sum_{l=1}^{N_{\rm L}}z_{l,t}\le N_{\rm T},\forall t,t=1,2,\cdots, N_{\rm TS}\\
			\text{}\hspace{15mm}\text{ C4:} \hspace{2mm} s_{u,l}\ge t^{(l)}_{\rm U} d_{u,l},\forall  u,u\in\mathcal{U}^{(l)},l\in L\\
			\text{}\hspace{15mm}\text{ C5:} \hspace{2mm} s_{l}\ge t_{\rm L} d_{l},l=1,2,\cdots,{N_{\rm L}}\\
			\text{}\hspace{15mm}\text{ C6:} \hspace{2mm}  z_{n_1,t} + z_{n_2,t} \leq 1, \forall (n_1, n_2)\in \mathcal{P} \\
			\text{}\hspace{16mm}\text{C7}: \left\{ \begin{array}{*{35}{l}}
				\text{N1}: \beta^{(l)}_{c,u}=0\text{ if }a^{(l)}_{c,u}=0\\
				\text{N2}: \beta^{(l)}_{c,u}> 0\text{ if }a^{(l)}_{c,u}=1\\
			\end{array}\right.
		\end{array}
	\end{equation}

	Clearly, this optimization problem in \eqref{main:ps2} is known as multi-objective mixed integer nonlinear programming (MINLP) because (i) there are multiple objective functions, (ii) constraints C4 and C5 are nonlinear due to multiplicative terms (multiplications between continuous variable and integer variable), and (iii) conditional constraints in \eqref{eqcc} are also nonlinear.  
	Since there usually exist multiple Pareto optimal solutions for multi-objective optimization problems, obtaining a solution for such  complicated systems in not a straightforward process.
	In this context, scalarizing functions can be used in solving multi-objective optimization problems based on different approaches \cite{Miettinen2002}. Specifically, scalarization mean converting the original problem with multiple objectives into a single-objective optimization problem. Then, if Pareto optimality of the resulted single-objective solutions can be guaranteed, then these solutions are Pareto optimal solutions to the original multi-objective problem.

	\section{Proposed Solution} \label{sec:proposed_solution}
	In the section, we elaborate the proposed solution to convert the multi-objective MINLP problem into a single objective MILP programming problem by scalarization of the multi-objective function and linearly approximating the nonlinear constraints.

	\subsection{Scalarization of the multiple objectives}
	Scalarizing multiple objectives can be done using different approaches. We can perform weighted sum linear scalarization with a weight of $1$ ($1$ unit on one objective is ``tradeable" for $1$ unit of any other objective) for each objective, thus by summing up all the objectives. Under such scalarization, the aggregation function is defined as $$\mathcal{F}(t)=t^{(l)}_{\rm U}+t^{(2)}_{\rm U}+\cdots+t^{(N_{\rm L})}_{\rm U}+t_{\rm L}$$ If there is a solution that maximizes all the objectives at once, then this choice of objective function is satisfying because this solution will be optimal for the weighted sum. However, such a solution is uncommon. Choosing the weighted sum as the scalarizing function limits the scope of potential solutions. Since there is no a certain preference on the trade-off between the two different objectives, trading one unit of $t_L$ for one unit on another objective seams reasonable in this case.
	
	For our demand-supply problem, we use a specific case of ordered weighted average based scalarizing function\cite{yager1988} by maximizing the minimum objective. The aggregating function $F$ is simply the minimum value across all objectives. Thereby, the aggregating function is given as 
	$$\mathcal{F}(t)=\min\left(t^{(1)}_{\rm U},t^{(2)}_{\rm U},\cdots,t^{(N_{\rm L})}_{\rm U},t_{\rm L}\right),$$ while the objective function is defined as $$\underset{a^{(l)}_{c,u},\beta^{(l)}_{c,u},z_{l,t}}{\max} \hspace{2mm} \mathcal{F}(t).$$ Since the target is to obtain a solution with a fair repartition across objectives, this approach is useful. In this formulation, users' demands are random values and the objective functions can take continuous values, and hence, the solver will not need to arbitrate between several solutions as the solutions more likely will not have the same minimum value. However, in case the solver need to arbitrate between several solutions with the same minimum value, we prefer the solution that dominates (e.g., more offered capacity) other than solutions on all objectives. In order to enable the solver to take this aspect into consideration in such circumstances, the aggregation function can be rewritten as
	$$\mathcal{F}(t)=\min\left(t^{(1)}_{\rm U},t^{(2)}_{\rm U},\cdots,t^{(N_{\rm L})}_{\rm U},t_{\rm L}\right)+\epsilon\left(\sum_{l=1}^{N_{\rm L}}t_{\rm U}^{(l)}+t_{L}\right)$$
	Now for the model, we need to add an extra variable denoted by $\theta$, and add the following $N_{\rm L}+1$ linear constraints:
	\begin{equation}
		\label{const1}
		\text{ C8}:\left\{ \begin{array}{*{35}{l}}
			\text{C8-a}:\theta \leq t_{\rm U}^{(l)}$ ,$l=1,\ldots,N_{\rm L}\\
			\text{C8-b}:\theta\le t_{\rm L}\\
		\end{array}\right.
	\end{equation}
	The objective function will then be $$\underset{a^{(l)}_{c,u},\beta^{(l)}_{c,u},z_{l,t},\theta}{\max}\left\{\theta +\epsilon\left(\sum_{l=1}^{N_{\rm L}}t_{\rm U}^{(l)}+t_{L}\right)\right\}.$$ Since $\theta$ is lower than all objectives, it is lower or equal to the objective with the minimum value, thus maximizing it yields the solution having the maximum minimum value.

	\subsection{Linearization of the products of two variables}
	As mentioned earlier the expression for supply capacity to each user is a a nonlinear function owing to the multiplicative term $\beta^{(l)}_{c,u}z_{l,t}$. In order to make it a linear function, new slack variables have to replace the previous terms as  $q^{(l)}_{c,u,t}=\beta^{(l)}_{c,u}z_{l,t}$. To linearize these nonlinear (multiplication of continuous variable and integer variable), we introduce the following linear constraints.
	\begin{equation}
		\label{const1}
		\text{ C9}:\left\{ \begin{array}{*{35}{l}}
			\text{C9-a}:q^{(l)}_{c,u,t}\ge 0\\
			\text{C9-b}:q^{(l)}_{c,u,t}\le z_{l,t}\\
			\text{C9-c}:q^{(l)}_{c,u,t}\le \beta^{(l)}_{c,u}\\
			\text{C9-d}:q^{(l)}_{c,u,t}\ge \beta^{(l)}_{c,u}-(1- z_{l,t})
		\end{array}\right.
	\end{equation}
	For the case, $z_{l,t}=0$, $q^{(l)}_{c,u,t}$ should be 0. The inequalities \{C9-a, C9-b\} causes $0\le q^{(l)}_{c,u,t}\le 0$, yielding $q^{(l)}_{c,u,t}$ to be 0. The other pair of linear constraints \{C9-c, C9-d\} returns $\beta^{(l)}_{c,u}-1\le q^{(l)}_{c,u,t} \le \beta^{(l)}_{c,u}$, and ${\lambda}_{c,u}=0$ conforms these inequalities. On the other hand, for the case $ z_{l,t}=1$, the product should be $q^{(l)}_{c,u,t}= \beta^{(l)}_{c,u}$. The inequalities \{C9-a, C9-b\} enforce $0 \le q^{(l)}_{c,u,t} \le 1$, which is satisfied by $q^{(l)}_{c,u,t}=\beta^{(l)}_{c,u}$. The second pair of inequalities \{C9-c, C9-d\} yields $\beta^{(l)}_{c,u} \le q^{(l)}_{c,u,t} \le \beta^{(l)}_{c,u}$, forcing ${\lambda}_{c,u}=f_{c,u}$ as needed. After the linear approximation, the supply capacity to user $u$ of cluster $l$, $s_{u,l}$ becomes an affine function of $q^{(l)}_{c,u,t}$, and thus, a linear function that can be written as
	
	\begin{equation}
		s_{u,l}=\sum_{t=1}^{N_{\rm TS}}\sum_{c=1}^{N^{(l)}_{\rm C}}q^{(l)}_{c,u,t} \mathcal{R}^{(l)}_{c,u}
	\end{equation}

	\subsection{Linearization of conditional constraints}
	Note that the {\textit{conditional}} constraints C7 in \eqref{eqcc} are nonlinear as well. The following linear constraints will linearize the nonlinear {\textit{conditional}} constraint.
	\begin{equation}
		\text{C7}:\left\{ \begin{array}{*{35}{l}}
			\text{L1}: \beta^{(l)}_{c,u}  \le Ma^{(l)}_{c,u}\\
			\text{L2}: \beta^{(l)}_{c,u}  \ge \epsilon - (1-a^{(l)}_{c,u})
		\end{array}\right.
	\end{equation}
	Here, $M$ is a large constant. $M$ is selected to be big enough so that the constraints do not influence the optimal solution/feasible set when the condition does not hold, i.e. in these models, it would be an upper bound on $a^{(l)}_{c,u}$. Therefore, we should set it equal to 1, which is the largest value that $\beta^{(l)}_{c,u}$ can reasonably take. Consequently, the linearized {\textit{conditional}} constraints eventually can be expressed as
	\begin{equation}\label{eq222}
		\text{C7}:\left\{ \begin{array}{*{35}{l}}
			\text{L1}: \beta^{(l)}_{c,u}  \le a^{(l)}_{c,u}\\
			\text{L2}: \beta^{(l)}_{c,u}  \ge \epsilon - (1-a^{(l)}_{c,u})
		\end{array}\right.
	\end{equation}
	Strict inequalities in N2 of \eqref{eqcc}, e.g., if $a^{(l)}_{c,u}=1,\text{ then }\beta^{(l)}_{c,u}>0$ is pointless in practical optimization, and thus, we come up with a significantly large $\epsilon$ that is non-zero and does not drown in the general precision and tolerance of numerical solvers, typically around $10^{-7}$. Accordingly, if $a^{(l)}_{c,u}=0$, then $\beta^{(l)}_{c,u}$ must equal $0$ by the first constraint while the second constraint has no effect. If $a^{(l)}_{c,u}=1$, then the second constraint prevails, and the first constraint has no effect, and delivers the expected mapping between the association indicator $a^{(l)}_{c,u}$ and the fill-rate variable $\beta^{(l)}_{c,u}$. Moving forward, after the scalarization of the multiple objectives and linear approximation of the non-linear constraints, the single objective MILP problem can be expressed as
	\begin{equation}
		\label{main:psfinal}
		\begin{array}{*{35}{l}}
			\underset{a^{(l)}_{c,u},\beta^{(l)}_{c,u},z_{l,t}}{\max}\hspace{5mm}\left\{\theta +\epsilon\left(\sum_{l=1}^{N_{\rm L}}t_{\rm U}^{(l)}+t_{L}\right)\right\}\\
			\text{}\text{subject to }\text{ C1:} \hspace{2mm}\sum_{\forall c, c\in \mathcal{C}^{(l)}} a^{(l)}_{c,u}\le \Delta_{\max},\forall  u,u\in\mathcal{U}^{(l)}, \vspace{1.5mm} \\
			\text{}\hspace{15mm}\text{ C2:} \hspace{2mm}\sum_{u=1}^{N^{(l)}_{\rm U}} \beta^{(l)}_{c,u}\le 1, \forall  c,c\in\mathcal{C}^{(l)}, \vspace{1.5mm} \\
			\text{}\hspace{15mm}\text{ C3:} \hspace{2mm}\sum_{l=1}^{N_{\rm L}}z_{l,t}\le N_{\rm T},\forall t,t=1,2,\cdots, N_{\rm TS}\\
			\text{}\hspace{15mm}\text{ C4:} \hspace{2mm} s_{u,l}\ge t^{(l)}_{\rm U} d_{u,l},\forall  u,u\in\mathcal{U}^{(l)}, l=1,2,\cdots,{N_{\rm L}}\\
			\text{}\hspace{15mm}\text{ C5:} \hspace{2mm} s_{l}\ge t_{\rm L} d_{l},l=1,2,\cdots,{N_{\rm L}}\\
			\text{}\hspace{15mm}\text{ C6:} \hspace{2mm}  z_{n_1,t} + z_{n_2,t} \leq 1, \forall (n_1, n_2)\in \mathcal{P} \text{ with } \mathcal{P}_{n_1 n_2}=1\\
			\text{}\hspace{15mm} \text{ C7}:\left\{ \begin{array}{*{35}{l}}
				\text{C7-a}: \beta^{(l)}_{c,u}  \le a^{(l)}_{c,u}\\
				\text{C7-b}: \beta^{(l)}_{c,u}  \ge \epsilon - (1-a^{(l)}_{c,u})
			\end{array}\right.\\
			\text{}\hspace{15mm} \text{ C8}:\left\{ \begin{array}{*{35}{l}}
				\text{C8-a}:\theta \leq t_{\rm U}^{(l)}$ ,$l=1,\ldots,N_{\rm L}\\
				\text{C8-b}:\theta\le t_{\rm L}\\
			\end{array}\right.\\
			\text{}\hspace{15mm} \text{ C9}:\left\{ \begin{array}{*{35}{l}}
				\text{C9-a}:q^{(l)}_{c,u,t}\ge 0\\
				\text{C9-b}:q^{(l)}_{c,u,t}\le z_{l,t}\\
				\text{C9-c}:q^{(l)}_{c,u,t}\le \beta^{(l)}_{c,u}\\
				\text{C9-d}:q^{(l)}_{c,u,t}\ge \beta^{(l)}_{c,u}-(1- z_{l,t})
			\end{array}\right.
		\end{array}
	\end{equation}
	which can solved using the MILP solver because it provides globally optimum and accurate solution \cite{Kumar2020}.

	\section{Simulation Results} \label{sec:simulation_results}
	In this section, the performance of the proposed joint BH-CA scheme to the multi-beam HTS systems is evaluated through simulations. The simulation setup includes 16-beam geostationary orbit (GEO) satellite that constructed to form 8 clusters from the total beam coverage pattern. The number of users in each beam is 12 that are randomly distributed over the coverage of the given cluster. Each beam has two carriers and the carrier bandwidth is $54$ MHz. The transmit power per beam is set to 12 Watt. To reflect the traffic demand variations in the system, 30\% of the users are having very high demands while the remaining users have low to average demands. The other  simulation parameters are provided in Table.~\ref{tab-123}.
	
	\begin{table}[h]
		\centering
		\caption{Simulation Parameters} 
		\label{tab-123}
		\begin{tabular}{|l |r|}
			\hline
			Parameter & Value \\
			\hline
			Power per transponder, $P_{\rm T}$ &\hspace{2mm}15 dBW\\
			System bandwidth, $B$ & \hspace{1.95mm}500 MHz \\ 
			Roll-off factor & \hspace{1.95mm}20\% \\ 
			Maximum number of carriers per user, $\Delta_{\max}$ & 2\\
			Duration of a time-slot, $T_{\rm slot}$ & \hspace{2mm}1.3 ms\\
			Hopping Window,  $T_{\rm H}$ & \hspace{1.95mm}64 $T_{\rm slot}$ \\ 
			Number of transponders, i.e., HPA, $N_{\rm HPA}$ & 8\\
			Polarization &   Dual\\
			\hline
		\end{tabular}
	\end{table}
	
	\begin{figure}
		\centering
		\includegraphics[scale=0.6500]{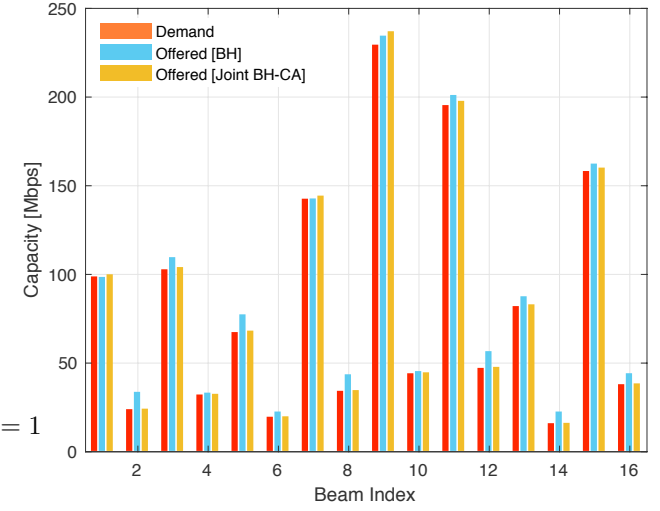}
		\caption{A comparison between the proposed joint BH-CA solution and the conventional BH scheme in terms of rate matching with the traffic demand under the same system setup and parameters. }
		\label{fig_BHCA_res1}
		\psfrag{Demand}{$Demand$}
	\end{figure}
	
	In Fig.~\ref{fig_BHCA_res1}, the proposed joint BH-CA scheme is evaluated in terms of its capability in matching the supply to the demand at the beam level and compared to the conventional BH scheme. It is evident from the bar-chart that both the proposed joint CA-BH solution and BH scheme perform equally well to a large extent in matching the supplies to the demands, and thus, efficiently utilizing the available satellite resources. The system has a total demand of 1325 Mbps while the total offered capacities by the conventional BH method and the proposed joint BH-CA scheme are 1409 Mbps and 1346 Mbps, respectively. This observation indicates that both schemes offer unused capacities to the system. However, the unused capacity using the BH solution is higher than that of the proposed BH-CA scheme. More importantly, our analysis reveals that user fairness performance when CA is employed with BH is better than the case of only BH is considered.

	\begin{figure*}[!t]
		\centering
		\includegraphics[scale=0.55]{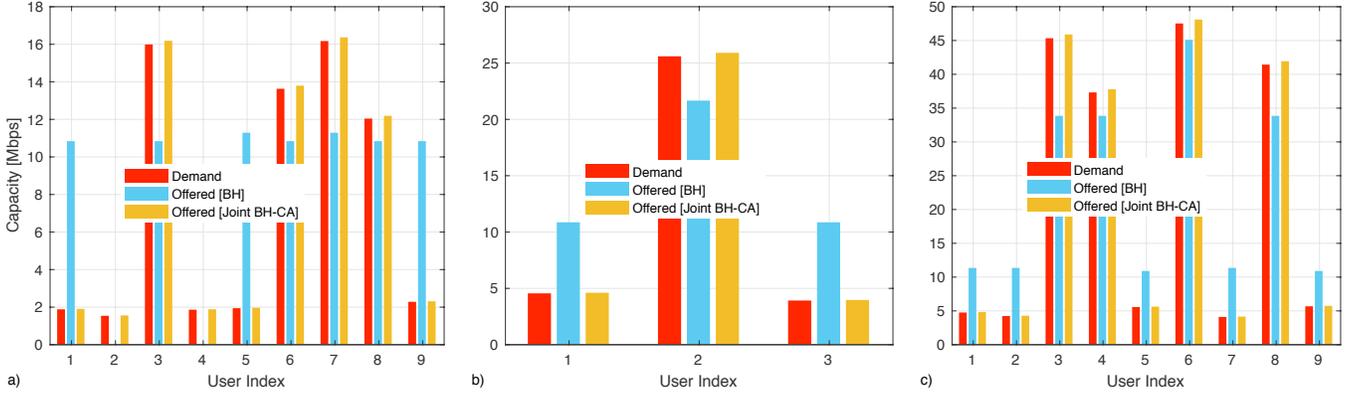}
		\caption{Rate matching performance with the proposed joint BH-CA solution and its comparison with equivalent BH system for three beams that have different traffic demands. a) First beam, b) Second beam, and c) Third beam. }
		\label{fig_BHCA_res2}
		\psfrag{Demand}{$Demand$}
	\end{figure*}
	
	\begin{figure}[!t]
		\centering
		\includegraphics[scale=0.6300]{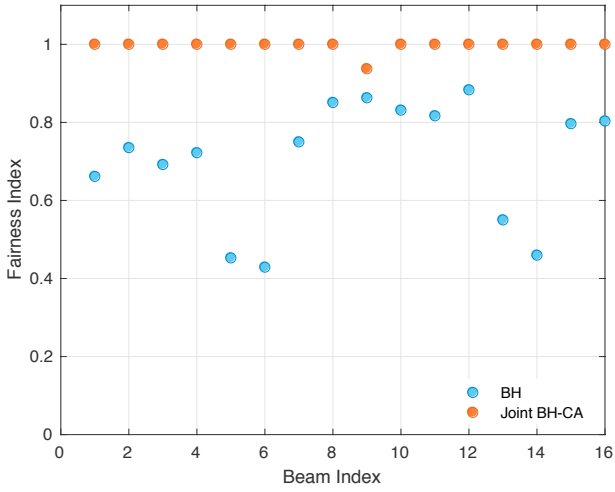}
		\caption{Fairness comparison between the proposed solution and conventional BH scheme at user-level. }
		\label{fig_BHCA_res3}
		\psfrag{Demand}{$Demand$}
	\end{figure}
	
	To examine user-fairness in the proposed solution comparing to the solutions with different objectives, the well-known Jain’s fairness index is a good tool to quantify the fairness measure. The Jain’s fairness index measures  how the provided rate matches the demand at a beam level as well as at a user level. Note that the relative allocation of the generic user $u$ in cluster $l$, $\zeta_u$ is the ratio between the offered capacity $s_{u,l}$ and the demanded/ideal capacity $d_{u,l}$, i.e., $\zeta_{u,l}=\frac{s_{u,l}}{d_{u,l}}$.
	In other words, we opt to obtain the relative allocations to be equal, i.e., $\zeta_{1,l}=\zeta_{2,l}=\cdots=\zeta_{N_{\rm U}^{(l)},l}$.
	We employ the Jain’s fairness index to measure the tightness of relative allocations at user level, which is defined as
	\begin{equation}
		\label{main:jfi}
		J_{{\rm FI},l}=\frac{\left(\sum\limits_{u=1}^{N_{\rm U}^{(l)}}\zeta_{u,l}\right)^{\!\!2}}{N_{\rm U}\sum\limits_{u=1}^{N_{\rm U}^{(l)}}\zeta_{u,l}^2}.
	\end{equation}
	A totally fair CA design will yield $J_{\rm FI}=1$. As the disparity increases, the fairness decreases. The $J_{\rm FI}$ is bounded between $\frac{1}{N_{\rm U}}$ and 1. Similarly, we can define fairness index at beam level.

	Next, the rate matching performance is evaluated at a user level in Fig.~\ref{fig_BHCA_res2} for both joint BH-CA and the conventional BH schemes. 
	The results are depicted for three different beams that have varying demands. For the BH scheme, we first solve the rate matching optimization problem when allowing only non-adjacent beams to be illuminated simultaneously. The number of assigned time-slots is then proportionally distributed among the users if the number of time-slots is higher than the number of users in the beam. Otherwise, if the number of assigned time-slots is equal or lower than the number of users, then each user will be assigned to a time-slot. 
	On one hand, although the conventional BH scheme delivers high rate-matching performance at the beam level, it fails to achieve the same performance at the user level unless there is only one user in each beam, which is not a practical scenario. 
	On the other hand, the rate-matching performance of the proposed BH-CA scheme at the user-level is as good as rate-matching performance at the beam-level. In our joint BH-CA scheme, the BH operation takes care of rate-matching at the beam level while the CA operation takes care of rate-matching at the user-level. Whereas, the unused capacity of the low demand users cannot be shared with the high demand users in the conventional BH scheme, and this is the key feature that makes the proposed BH-CA scheme outperforms the  conventional BH approach.

	Fig.~\ref{fig_BHCA_res3} investigates the Jain's fairness index for both the proposed joint BH-CA solution and conventional BH scheme. The fairness factor is an important metric to measure the rate-matching performance with respect to the traffic demand. 
	The Jain's fairness index values for all beams with the joint BH-CA scheme are very  high and mostly reaching the optimal value that is one. Whilst, the conventional BH solution fails to provide high and uniform fairness index values across the beams in the system. 
	Therefore, the proposed joint BH-CA scheme is a very effective in reducing the unused capacity while ensuring high user-fairness and beam-fairness over the entire system. Additionally, this also implies that  the unsatisfied traffic demand is lower with the joint BH-CA scheme than the BH without CA, particularly in the presence of heterogeneous traffic demand.

	\section{Conclusions} \label{sec:conclusions}
	In this paper, a novel joint BH-CA scheme is developed for the multi-beam HTS systems. The high heterogeneity of satellite traffic demands in terms of time and location as well as the various QoS profiles of users have inspired this combination to harness the unique synergetic interaction between BH and CA in satisfying the spatiotemporal traffic variations and improving system performance.  It has been shown in this paper that joint BH-CA scheme exploits the symbiotic convergence between BH that offers flexible satellite resource utilization in the time domain and CA scheme that increases the resource allocation flexibility in the frequency domain. To this end, a multi-objective mixed integer nonlinear programming (MINLP) optimization problem has been formulated and solved to realize the proposed concept. The performance of the proposed BH-CA approach has been evaluated comparing to the conventional BH scheme without CA. The conducted analyses and evaluations have clearly verified the superiority of the proposed BH-CA scheme over the conventional BH in fully satisfying the requested demand and ensuring the QoS fairness. The proposed joint BH-CA provides an excellent rate matching performance not only at a beam level but also at a user level, which validates its feasibility and efficiency for the multi-beam HTS systems. 

\linespread{1}
\bibliographystyle{IEEEtran}
\bibliography{IEEEabrv,References}
	
\end{document}